\documentclass[letterpaper, 10 pt, conference]{ieeeconf}
\IEEEoverridecommandlockouts    
\usepackage{times}
\usepackage{color}
\usepackage{graphicx}
\usepackage{hyperref}
\usepackage{latexsym}
\usepackage[caption=false]{subfig}

\newcounter{question}
\newcommand{\question}[1]{Q\refstepcounter{question}\thequestion: #1}

\title{
SoK: A Survey of Open-Source Threat Emulators
}

\author{Polina Zilberman$^{1}$ and Rami Puzis$^{1}$ and Sunders Bruskin$^{1}$ and Shai Shwarz$^{1}$ Yuval Elovici$^{1}$
\thanks{$^{1}$Department of Software and Information Systems Engineering and Cyber@BGU, Ben-Gurion University of the Negev, P.O.B. 653 Beer-Sheva, Israel.
        {\tt\small \{sundersb\}@post.bgu.ac.il},
        {\tt\small \{polinaz,puzis,elovici\}@bgu.ac.il}}%
}

\begin{document}

\maketitle
\thispagestyle{empty}
\pagestyle{empty}

\begin{abstract}
Threat emulators are tools or sets of scripts that emulate cyber attacks or malicious behavior. 
They can be used to create and launch single procedure attacks and multi-step attacks; the resulting attacks may be known or unknown cyber attacks
The motivation for using threat emulators varies and includes the need to perform automated security audits in organizations or reduce the size of red teams in order to lower pen testing costs; or the desire to create baseline tests for security tools under development or supply pen testers with another tool in their arsenal.

In this paper, we review and compare various open-source threat emulators. 
We focus on tactics and techniques from the MITRE ATT\&CK Enterprise matrix and determine whether they can be performed and tested with the emulators.
We develop a comprehensive methodology for our qualitative and quantitative comparison of threat emulators with respect to general features, such as prerequisites, attack definition, clean up, and more. 
Finally, we discuss the circumstances in which one threat emulator is preferred over another.

This survey can help security teams, security developers, and product deployment teams examine their network environment or products with the most suitable threat emulator. 
Using the guidelines provided, a team can select the threat emulator that best meets their needs without evaluating all of them.

\end{abstract}

\section{Introduction}
\label{sec:intro}
It is essential for IT security professionals to identify weaknesses in security systems before cyber criminals exploit them. 
Red teams are groups of white-hat hackers that perform pen testing by assuming an adversarial role. 
In addition to finding unpatched vulnerabilities, one of the most important objectives of a red team is evaluating the organization's security readiness, active controls, and countermeasures by emulating a full attack lifecycle. 
Extensive what-if analysis is necessary to evaluate an organization’s response to an attack that may penetrate its premises~\cite{Rajendran2011,Muehlberghuber2013,Applebaum2016,rege2019need,Rastegari2013,MANSFIELDDEVINE20188}. 


Threat emulation platforms, such as Red-Team Automation (RTA)~\cite{rta} and Atomic Red Team~\cite{atomic-rt},
significantly accelerate and simplify what-if analysis in diverse scenarios.
Threat emulators also make it easier for IT professionals without red team qualifications to test an organization's security controls and countermeasures.
\emph{A good threat emulator makes it is easy to challenge the organization's security controls with a wide variety of realistic multi-step attacks in a controlled manner.} 
IT Professionals cannot afford to explore and use multiple threat emulators. 
Thus, there is a need to review existing emulators and develop a methodology to support threat emulator selection; doing so will assist IT professionals in choosing the threat emulator that best meets organizational needs. 
To the best of our knowledge, no rigorous comparison of threat emulators has been published.

In this paper, we survey general purpose open-source threat emulators suitable for post-compromise analysis, excluding tools designed for specific types of systems. 
For example, security assessment tools such as sqlmap~\cite{sqlmap} and w3af~\cite{w3af}, respectively Web application attack and audit frameworks, are not included in this survey, because they are too specific and are not general purpose threat emulators.
Various fuzzing tools~\cite{Bohme2017,Wang2010,Holler2012}, as well as network and vulnerability scanners such as Nmap~\cite{Lyon2009} or Nessus~\cite{rogers2011nessus}, are also excluded from this survey, because they are not used to emulate complex multi-step attack scenarios.


The threat emulators are reviewed and compared with respect to tactics and techniques from the MITRE ATT\&CK matrix and criteria such as prerequisites, attack scenario definition, and clean up. 
The contributions of this study are fourfold:
\begin{enumerate}
    \item We present well-defined criteria and a 
    methodology for the evaluation and comparison of threat emulators.
    \item We present a thorough review of the following threat emulators: APTSimulator, Red Team Automation, Uber's Metta, CALDERA, Atomic Red Team, Infection Monkey, PowerSploit, DumpsterFire, Metasploit, BT3 and Invoke-Adversary.
    \item  We present a taxonomy that illustrates the qualities of the threat emulators.
    \item  We provide actionable guidelines for choosing the best threat emulator given a specific organizational environment and a set of security assessment tasks.
\end{enumerate}


The rest of the paper is structured as follows. 
Section~\ref{sec:mitre-table} provides background on the attack capabilities matrix. 
In Section~\ref{sec:ecosystem}, we define the threat emulation ecosystem. Section~\ref{sec:criteria} presents the criteria used to review and compare the threat emulators. 
A comprehensive methodology for evaluating threat emulators is presented in Section~\ref{sec:emulators}, followed by highlights of our review of the 11 threat emulators. 
Next, in Section~\ref{sec:taxonomies}, we introduce a comprehensive emulator taxonomy and guidelines for threat emulator selection. 
Section~\ref{sec:summary} summarizes the paper.

\section{Background: MITRE ATT\&CK}
\label{sec:mitre-table}
MITRE's ATT\&CK is an open knowledge base of adversarial tactics and techniques which is continuously updated based on real-world observations~\cite{strom2017finding}.
MITRE's ATT\&CK is considered state-of-the-art when it comes to describing common behavior of adversaries.
Many companies, such as D3 Security and AlienVault, rely on this knowledge base in threat intelligence that they consume or provide.

Many threat emulators are designed with respect to the techniques listed in the MITRE ATT\&CK Enterprise matrix\footnote{\url{https://attack.mitre.org/techniques/enterprise/}}. 
Threat emulators implement a variety of attack scenarios in which different tactics and techniques are employed. 
The larger are the number of tactics supported by an emulator and the number of techniques achieving each tactic, the higher is the diversity of the attack scenarios supported by an emulator.   
Diversity of supported attack scenarios aids the red team in discovering security flaws. 
MITRE ATT\&CK matrix helps systematically evaluating the diversity of attack scenarios that can be created by an emulator.

In this paper, the threat emulators are reviewed and compared with respect to tactics and techniques from the MITRE ATT\&CK matrix and criteria, we introduce the MITRE terminology used in the rest of this paper. 
\emph{Tactics} represent the adversary's technical goals, such as achieving persistence, privilege escalation, defense evasion, and more.
An attack's lifecycle, also referred to as the kill chain~\cite{hutchins2011intelligence,pols2017,case2016analysis}, is typically a series of goals achieved during the course of the attack.  Every tactic includes a variety of \emph{techniques}, also known as attack patterns~\cite{barnum2008common}, which achieve similar goals. 
A technique may achieve multiple goals, thus being associated with a number of tactics. 
Techniques are abstract descriptions that distill the essence of many similar attack implementations.
The specific implementations of the techniques are referred to as \emph{procedures}. 
For example, in order to achieve persistence (a tactic), an adversary creates a scheduled task (a technique) using the following Windows command: \texttt{schtasks /create /sc /daily /tn <folder path>\textbackslash<task name> /tr} (a procedure).



A brief explanation of the tactics listed in MITRE's ATT\&CK matrix and their contribution to the success of an attack is provided below. 
These tactics will be used as attributes in the detailed threat emulator review in Section~\ref{sec:emulators}.

\emph{Initial Access} tactic includes all of the techniques that enable malware to infiltrate a machine or a network environment. 
Pen testing focuses on this tactic.

\emph{Execution} tactic includes all of the techniques that result in the execution of attacker controlled code on a victim's machine. 
The more \emph{execution} techniques the attacker has in his/her toolbox, the more obstacles, such as the PowerShell execution policy, he/she can overcome.

\emph{Persistence} assures that even if the compromised machine restarts, the malware will persist. 
The fact that initial access is risky for the adversary and requires significant effort makes persistence crucial in the attack's lifecycle.


\emph{Privilege escalation} is required in order to access important assets only accessible with superuser permissions. 
Superuser (or administrative) permissions also facilitate the achievement of other goals, such as defense evasion.

\emph{Defense evasion} 
includes techniques that enable the attack to stay under the radar of security defense tools.

\emph{Credential access} is a goal pursued by most attackers. 
Having multiple user credentials facilitates privilege escalation, persistence, and defense evasion. 
For example, specific user credentials may be required to access a key system.

\emph{Discovery} 
includes techniques used by malware to discover assets, network topology, security controls, vulnerabilities, etc.

\emph{Lateral movement} includes techniques are used to reach the target assets spreading out from the initial entry point.
Emulating lateral movement is a crucial threat emulator capability when evaluating network security systems.

\emph{Collection} includes techniques for collecting the target data assets. 
In addition to espionage and intelligence, collection techniques also facilitate defense evasion and credential access tactics.

\emph{Exfiltration} includes techniques for transferring the collected data from the compromised environment to the attacker.

\emph{Command and control} includes techniques that enable the attackers' interactions with their tools installed on the compromised network.

\section{Threat Emulation Ecosystem}
\label{sec:ecosystem}

\begin{figure*}
    \centering
    \includegraphics[width=\textwidth]{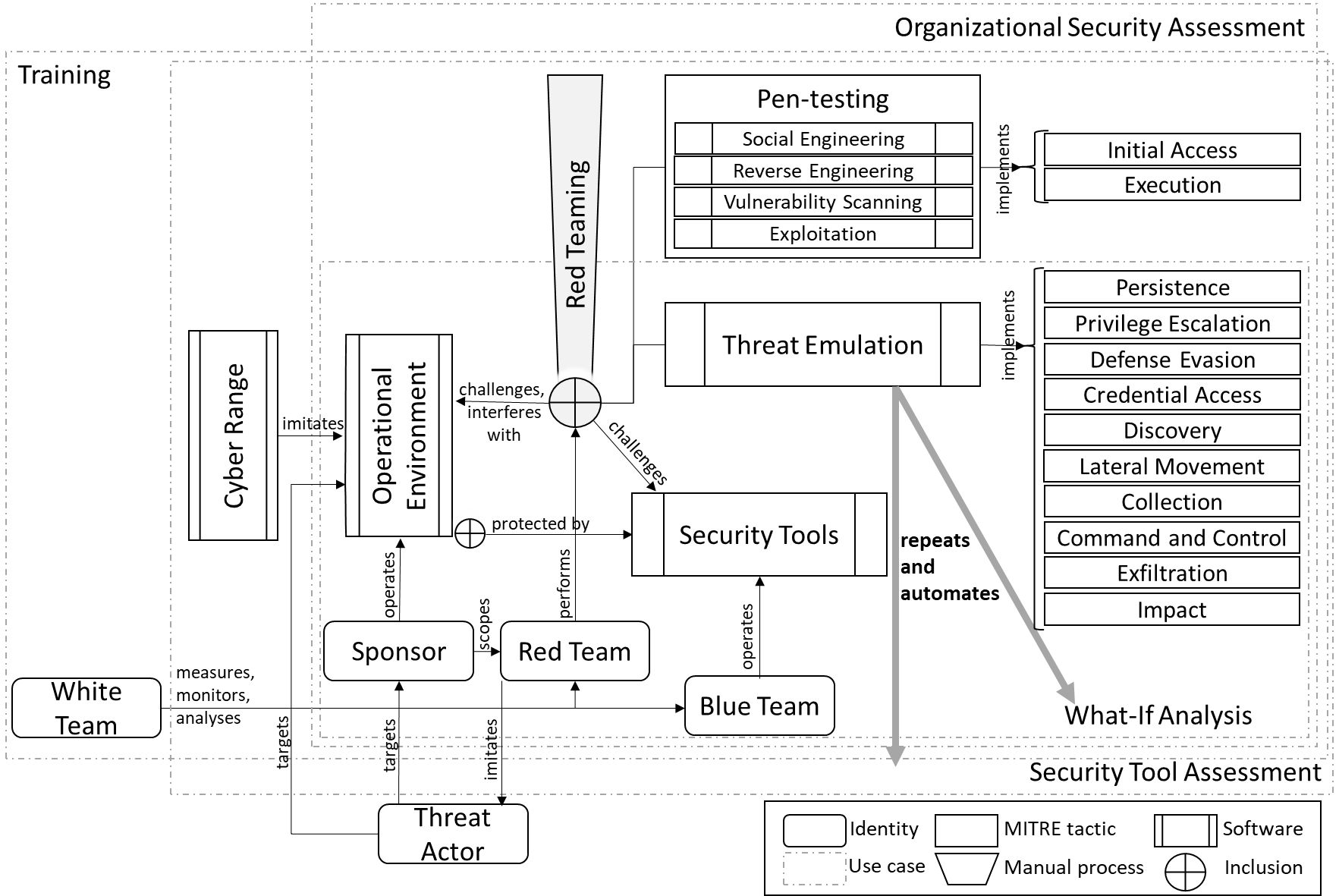}
    \caption{The threat emulation ecosystem}
    \label{fig:ecosystem}
    \vspace{-0.5cm}
\end{figure*}

Threat emulators are an integral component of comprehensive red teaming. In this section, we define the threat emulation ecosystem (see Fig.~\ref{fig:ecosystem}), including the stakeholders and relevant software processes, and their relationships in the threat emulation ecosystem. 
We map the MITRE tactics (see Section~\ref{sec:mitre-table}) to the red teaming software components. 
The ecosystem elements are framed with respect to four threat emulation use cases: training, security tool assessment, organizational security assessment, and what-if analysis.

\subsection{Stakeholders} 

\noindent The \emph{sponsor} is the organization that employs the red team to perform a security assessment.

\noindent The \emph{threat actor} is the adversarial entity that targets the sponsor. 
The threat actor's motivations may vary depending on the sponsor's assets. His/her attacks can compromise the sponsor's operational environment. Information about the threat actors that are most relevant to the assessed organization will enable the red team to take on the role of the attacker~\cite{Bishop2007,engebretson2013basics} more realistically and mimic the behavior of the threat actors~\cite{mejia2008red,oakley2019,MANSFIELDDEVINE20188}. 

\noindent The \emph{red team} is hired to challenge the security of an organization and may include pen testers, social engineers, reverse engineers, and intrusion detection specialists; in rare cases, the team may also include language specialists~\cite{adkins2013red}. 
The sponsor and red team agree on the assessment target and scope in light of the organization's critical assets and possible attack vectors~\cite{Bishop2007,oakley2019}. 
The scope is also limited by ethical considerations and legislation~\cite{MANSFIELDDEVINE20188}.


\noindent The \emph{blue team} defends the organization's systems by deploying and using security tools to detect and mitigate the activities of the threat actors.

\noindent The \emph{white team} is responsible for the coordination, execution, and analysis of the red team-blue team exercises~\cite{Mirkovic2008,Henshel2016}. 
White teams are most important during training.

\subsection{Software and Tools}

\noindent The \emph{operational environment} is the hardware and software that supports the organization's operation. 
The security needs arise from the type of assets within the operational environment, the threat actors, and the regulations the organization is subject to~\cite{work2019wolf}. 
The operational environment also affects the type of security assessment and training that can be performed in the operational environment itself and the type of training and assessment that must be performed in a non-operational environment, such as a cyber range~\cite{mejia2008red}.


\noindent\emph{Cyber range} is a term used for a virtual environment created for cyber security training. 
Some operational environments are too critical to risk malfunction during training or security assessment. 
In such cases, a safe environment that imitates the real operational environment is required~\cite{mejia2008red,oakley2019}. 
Such operational environments include critical infrastructure, e.g., supervisory control and data acquisition (SCADA) systems, where a single glitch resulting from red team activities could lead to a disastrous malfunction. 
Another reason for using a cyber range is the fact that threat emulation can trigger false alarms which are hard to distinguish from real attacks~\cite{work2019wolf}.

\noindent\emph{Security tools} represent the collection of intrusion detection, antivirus, firewall, SIEM (security information and event management), EDR (endpoint detection and response), and other systems deployed within the operational environment to protect the organizational assets. 
Security tools are operated by the blue team.

\noindent\emph{Vulnerability scanners, reverse engineering tools, exploits, fuzzers, etc.} are part of the red team's toolset. 
They are used by red teams to actively probe the operational environment and security tools for weaknesses and vulnerabilities. 
Following Applebaum et al.~\cite{Applebaum2016} and Adkins~\cite{adkins2013red}, we consider all tools used to identify opportunities for unsolicited access or code execution, including social engineering, as pen testing.



Today, cyber security defense paradigms assume that the organization is already compromised~\cite{gault2015}. 
Consequently, perimeter defense is insufficient, and the defenses deployed need to identify and mitigate malicious activities throughout the entire attack lifecycle. 
Security assessment of the organization in post-compromise scenarios is critical. 
Repeated assessment using red teams is expensive due to the expertise required. 
Manual red teaming also lacks the ability to replicate the security assessment in a fully repeatable manner.

\emph{Threat emulators}, the software tools that are the main focus of this survey, aid in the automation of post-compromise red team activities~\cite{Applebaum2016}, compensating for any expertise the red team may lack. 
Although, human white-hat (ethical) hackers cannot be replaced by automatic red teaming tools, the use of threat emulators can increase the productivity of red teams or reduce the level of expertise required of the red team~\cite{oakley2019}. 
Challenging the security array in many different scenarios rapidly and repeatedly is especially important during training and what-if analysis. When emulating attack techniques that may endanger the operational environment or personnel (e.g., persistence, credential access, lateral movement, collection, impact), threat emulators usually operate by leaving forensic evidence that mimics the execution of specific attack procedures. Emulators can execute real attack procedures without causing actual damage or compromising the organization.

Vulnerability scanners and threat emulators can interfere with the network and systems in place. 
Therefore, they should be used with caution within the operational environment. 
The extent to which pen testing tools are employed (frequency, intensity, volume, etc.) is defined by the sponsor and red team within the scope of red teaming.

Red team-blue team exercises should be used in a dedicated environment, like a cyber range, where there is no interference in the operational environment. 
The cyber range should imitate the operational environment to the extent possible in order to increase the suitability of the security tools and their operators (the blue team) to the organization.

\subsection{Red Teaming Use Cases}
\noindent\emph{Training.}
The terms red team and blue team originate from military exercises where a blue team is responsible for protecting an asset, and the red team's role is to challenge the blue team's plan of defense iteratively~\cite{mejia2008red}. 
The result of training should be a comprehensive understanding of the effectiveness of the asset's security mechanisms/protocols, as well as improved response in terms of mitigation and efficiency. 
In some cases, a white team oversees the training exercise(s), coordinating between teams and analyzing their performance. 
Often the training takes place in a controlled environment like a cyber range.

Threat emulators can be used to train the blue teams and test their capabilities in handling security events. 
Emulators can create predefined and well-designed scenarios that provide full coverage of essential detection and response cases. 
In the absence of a qualified red team, an operator without extensive security expertise can use a threat emulator to train the blue team members in typical scenarios.

\noindent\emph{Security Tool Assessment}. 
Such assessment and evaluation is required to correctly position a security product within the cyber security landscape. 
Organizations' security needs can vary. 
The security requirements may depend on the type of assets an organization has, the regulation it is subject to, the organization's size, and other characteristics of the operational environment.

Security tool assessment encompasses two perspectives: (1) assessing the contribution of a tool to the security preparedness of the organization, and (2) testing the resilience of a tool (or a new version of it), also known as subversive exploitation~\cite{oakley2019,MANSFIELDDEVINE20188}.
Assessment of security products should never compromise an organization~\cite{Mirkovic2008}. 
Therefore it is preferable to evaluate a security tool within a cyber range that mimics the environment of the target organization. Alternatively, an organization can use threat emulators to perform comparisons and assessments of security tools without compromising the organization's security.

The ability to create and reproduce simulated attacks helps when comparing and contrasting security. 
MITRE ATT\&CK Evaluation\footnote{https://attackevals.mitre.org/} is an example of a service that assesses the capabilities of different types of security tools. 
In particular, it specifies the adversarial techniques that each tool evaluated can mitigate.



\noindent\emph{Organizational Security Assessment.}
Security regulations and standards, such as the EU General Data Protection Regulation (GDPR)~\cite{voigt2017eu}, ISO/27001, or the NIST catalog of security and privacy controls~\cite{force2013security}, define a set of security controls as well as standard security assessment activities. 
For example, section CA-8 in NIST Special Publication 800-53 on security controls defines the need for independent security assessment using red teams and red team exercises.

Due to the constant development of the cyber threat landscape, organizations must routinely review, document, and update all aspects of security. 
The red team's role is to identify weak points and security breaches through interaction with different planes of the organization: systems, users, applications, etc.~\cite{oakley2019,MANSFIELDDEVINE20188}.
In-house or outsourced red teams assess the organization's security by simulating attacks that do not compromise the security but thoroughly test it~\cite{radack2008guide}. 
Launching a wide range of emulated attacks allows the organization to better understand its vulnerabilities and weaknesses; as a result, the organization can take measures to improve security.

\noindent\emph{What-If Analysis}
This type of analysis assesses potential damage under different threat assumptions. 
What-if analysis disregards why and how a change in the operational environment has occurred and focuses on the consequences. 
Such analysis is important for evaluating cyber defenses' robustness to changes in both the course of the attack (red team perspective) and the operational environment (blue team perspective)~\cite{baiardi2019avoiding}. 
Assessment of the potential impact of security events using what-if analysis helps prioritize the respective mitigation activities.

By using threat emulators, the security team can automate large parts of the what-if analysis process. 
Threat emulators allow the reproduction of attack scenarios while only changing specific attack parameters.

\section{Criteria for Comparison}
\label{sec:criteria}

\begin{figure}
\centering
\includegraphics[width=1.0\columnwidth]{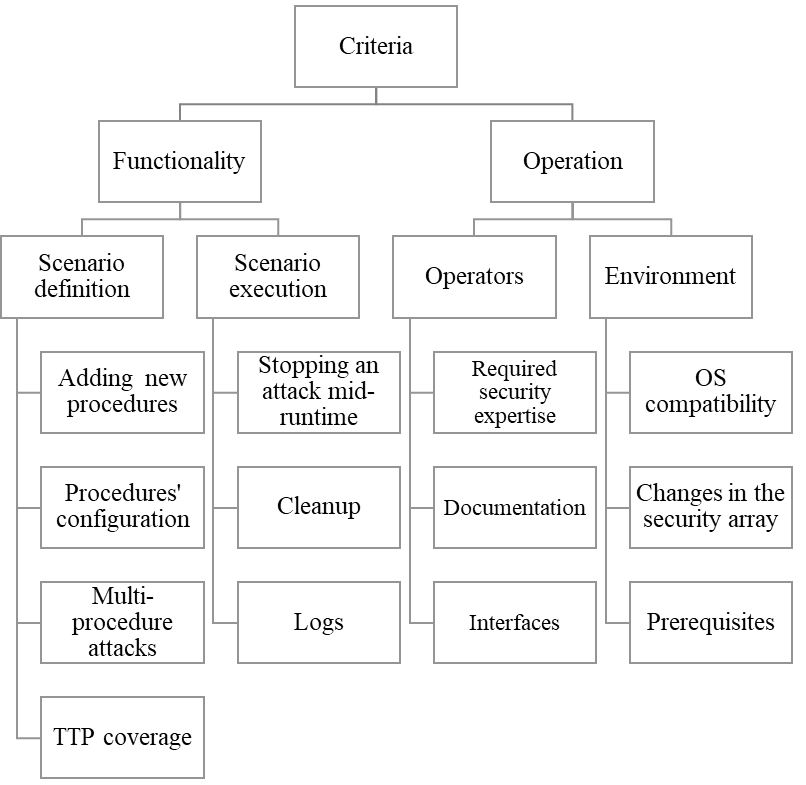}
\caption{\vspace{-0.2cm}
Criteria hierarchy for comparing threat emulators}
\label{fig:criteria}
\vspace{-0.5cm}
\end{figure}

One of the objectives of threat emulation is to accelerate and simplify the what-if analysis of organizational security controls and countermeasures in diverse scenarios. 
Suitability of a threat emulation framework depends on the capabilities of the red/blue teams, as well as on various organization constraints. 
This section presents criteria for comparing and evaluating threat emulators (see Fig.~\ref{fig:criteria}).



\subsection{Environment}
Determining whether or not the emulator can operate in the target environment is the first and most basic criteria for assessing the emulator's compatibility. 
This includes operating system compatibility, required changes in the security array, special privileges, etc.

\subsubsection{Operating system compatibility}
Usually organizations work with different operation systems (OSs). 
\emph{Ideally a threat emulator should support all operation systems used by the organizational endpoints.} 
We do not include requirements of the CNC component of the emulator in the OS compatibility criterion, because it is not part of the organizational environment tested.

The OS compatibility criterion may include one or more of the following OSs: Windows, Linux, MacOS, Android, iOS. 
Due to the ubiquitous nature of mobile devices, it is very important to include them in the attack scenarios tested. 
Thus, we include mobile platforms in this threat emulator evaluation criterion, although, none of the emulators reviewed support them.

\subsubsection{Changes in the security array}
\label{sec:level-of-detect}
Every organization has various security tools that provide some level of protection. 
Antivirus (AV) software and firewalls (FWs) are elements of the organizational security array. 
Some emulators require disabling AV and FWs to operate successfully. 
For example, a threat emulator that relies on remote administration tools (RATs) may require disabling the real-time antivirus protection in order to operate successfully. 
Such an emulator is not suitable for evaluating antivirus protection effectiveness.

\emph{A good threat emulator should operate without requiring changes in the organizational security array.} 
In the ideal case, the security controls and countermeasures evaluated are capable of detecting the emulated attacks but not the emulator's agent.
The value of this criterion represents the set of security controls that must be disabled in order to execute the emulator. 
There are three possible values of this criterion: 
\newline\noindent $\emptyset$ -- The emulator can operate successfully without disabling any security controls. 
\newline\noindent AV -- Antivirus software should be disabled.
\newline\noindent FW -- The firewall should be disabled.

\subsubsection{Prerequisites}
This criterion specifies whether there are any special prerequisites for the threat emulator to work. 
It may include third party software tools (such as mimikatz\footnote{\url{https://github.com/gentilkiwi/mimikatz}, \url{https://www.varonis.com/blog/what-is-mimikatz/}}), hardware components (such as a dedicated CNC server), special privileges (if the emulator's agent requires superuser privileges), etc. 
\emph{Similar to any redistributed package, a good threat emulator should be self-contained.}
The value of this criterion can consist of one or more of the following:
\newline\noindent $\emptyset$ -- The emulator is self-contained and does not require special privileges. 
\newline\noindent 3rd -- The emulator requires the installation of third party tools on the endpoints. 
\newline\noindent Priv -- The emulator requires special privileges (e.g., superuser or running in a kernel mode).


We consider an emulator to be self-contained if it includes everything that is required to operate on the endpoints. 
We exclude any prerequisites for the dedicated servers from this criterion, because their impact on the deployment of the emulator is minimal compared to any required installations on the endpoints. 
In addition, since we focus on open-source threat emulators, we assume that all prerequisites are open-source as well.

\vspace{-0.2cm}
\subsection{Scenario Definition}
\label{sec:attack-compl}
When evaluating security solutions it is often necessary to mimic complex and sophisticated real attacks. 
A threat emulator should be able to define all steps of the emulated attacks and produce realistic artifacts. 
Mimicking sophisticated real attacks includes both the ability of the emulator to leave detectable traces, as well as its ability to cover its tracks when emulating defense evasion techniques.

\subsubsection{Adding new procedures}
Cyber criminals constantly develop new exploits and perform unknown attacks. 
Hence, \emph{the ability to add new procedures is crucial for a future-proof threat emulator.} 
A good emulator should be able to handle the addition of new procedures for techniques it already implements as well as procedures that introduce new techniques. 
This criterion is binary: 
\newline\noindent Yes -- The emulator supports adding new custom procedures.   
\newline\noindent No -- The emulator does not provide interfaces or infrastructure for adding new custom procedures. 


\subsubsection{Procedures' configuration}
Different organizations have different assets, network structures, and security controls. 
Consequently, it is important to be able to easily modify the operation of an emulated attack in order to fit the organization. 
For example, one should be able to configure the IP range, ports, and protocols of a procedure that implements network scanning.

On the one hand, if the emulator does not support the addition of new procedures or the configuration of built-in procedures, its applicability is very limited. 
On the other hand, adding a new procedure instead of configuring an existing procedure is extremely ineffective. 
Moreover, the ability to configure procedures is required when emulating a large number of attack variants in a row. 
\emph{A good threat emulation framework should have the capacity to configure both built-in procedures and new custom procedures.} 
For example, an emulator that executes remote code on an endpoint should at least ensure that the parameters of the executed procedures are set before transferring the code to the endpoint.
We consider three possible values for the configurability of the threat emulator:
\newline\noindent Low -- The emulator doesn't support repeated execution of the same attack scenario with different parameters. 
\newline\noindent Med. -- The emulator does support repeated execution of the same attack scenario with different parameters.
\newline\noindent High -- Like the medium value, but the emulator also has the ability to easily configure new procedures.   

We assign a low configurability value to emulators with hardcoded procedure parameters or when procedures can only be configured manually, e.g., through a GUI. 
If the emulator supports configuration files or can be configured through a CLI or API, it is possible to automate the execution of attack scenarios with different parameters.



\subsubsection{Multi-procedure attacks}

An attack can be referred to as a set of procedures launched along a specific timeline. 
In order to emulate a realistic attack, it is necessary to combine multiple procedures together. 
\emph{Threat emulators should both provide multiple built-in attack scenarios and support the creation of new multi-procedure attacks.} Built-in attack scenarios may assist red teams that are rapidly testing new security solutions, IT professionals who are not qualified red team practitioners, and novice security personnel undergoing training. 
The ability to create custom attack scenarios that recombine tactics, techniques, and procedures is very important for challenging the organizational security array and performing diverse what-if analysis.
This criterion may include one or more of the following values: 
\newline\noindent $\emptyset$ -- No support of multi-procedure attacks. 
\newline\noindent Built-in -- Includes ready to run multi-procedure attacks.
\newline\noindent Custom -- Multi-procedure attacks can be added.


\subsubsection{Tactics, techniques, and procedures (TTP) coverage}

The diversity of a threat emulator is derived from the assortment of attack techniques it can emulate. 
To map the capabilities of a threat emulator, we refer to the set of tactics, techniques, and procedures in the MITRE ATT\&CK knowledge base. The more tactics a threat emulator covers, the more complete with respect to an attack's lifecycle the emulated attacks can be. 
The more techniques a threat emulator implements, the more comprehensive the resulting assessment is. 
In an ideal case \emph{a threat emulator should contain multiple procedures implementing every known adversarial technique.}

At least one procedure/technique is required to claim that a technique/tactic (respectively) is supported by an emulator. 
We define the following three levels of TTP coverage based on the tactics and techniques supported:
\newline\noindent Low -- The number of supported tactics is less than six or the number of supported techniques is less than 20. 
\newline\noindent Med. -- The emulator supports six or more tactics and 20 or more techniques.  
\newline\noindent High -- The emulator supports more than eight tactics and over 40 techniques.

The thresholds above are quite low given the 12 tactics and 266 techniques currently listed by MITRE in the Enterprise ATT\&CK matrix. 
Unfortunately, the overall TTP coverage provided by current threat emulators is very low. 
The thresholds were set such that they partition the set of emulators into three non-trivial groups.

\vspace{-0.1cm}
\subsection{Scenario Execution}
\subsubsection{Stopping an attack mid-runtime}
When examining an operational environment during the execution of a simulated attack, an operator may want to stop the attack for various reasons. 
For example, a simulated attack might consume too many computational resources or otherwise harm the system's operation, or the endpoints may require crucial updates. 
There may also be a need to change the attack scenario after the evaluation process has started. 
For example, the red team may find that the scenario is missing a crucial component required for an effective assessment, a realization that would necessitate stopping and restarting the evaluation. 
Consequently, \emph{it is important that the emulator has the ability to stop a simulated attack at any stage before it finishes running.}
Note that changing an attack scenario while it is being executed results in a new attack scenario that is different from both the new one and the old one. 
Thus, we do not consider it a useful capability. 
This criterion is binary: 
\newline\noindent Yes -- The emulator supports stopping an attack scenario while it is running without far-reaching consequences.   
\newline\noindent No -- The emulator does not provide such functionality, and the operator must manually stop all of the attack components.


\subsubsection{Cleanup}
After an emulated attack scenario has been completed or stopped in the middle, \emph{the machines must be rolled back to their previous state, and all traces of the attack must be cleaned up and removed.} 
This functionality is especially important when emulating a large number of attacks. 
Cleanup may include, for example, changing registry values, files, etc.

During cleanup a threat emulator should remove artifacts that it has created. 
However, send and forget actions performed by the emulator, such as pinging a remote server, may be logged by the monitoring system. 
As a result, an alert may be created and stored within the SIEM. 
An emulator is unaware of the security, logging, or monitoring tools in the network. 
In cases in which threat emulation is performed on a dedicated cyber range, it is possible to roll the machinery back to its previous state and emulate additional attacks.

However, in an operational network, reverting the response of the security tools to the activity of the threat emulator is not a reasonable requirement. 
This will create too strong coupling between the red and blue teams' instrumentation. 
Nevertheless, the operational staff should be aware of the possible consequences of the threat emulation process, including possible collateral damage and false alarms~\cite{work2019wolf}.
This criterion is ternary: 
\newline\noindent No -- The emulator does not support cleanup. 
\newline\noindent Proc -- Cleanup of individual procedures is supported.   
\newline\noindent Attk -- Cleanup of multi-procedure attacks is supported.


Although, none of the emulators reviewed facilitate cleanup for new custom procedures, it is important to encourage such implementation at the API level. 
The teardown functions in unit tests are a good example of cleanup facilitated by a framework. 
In the case of multi-procedure attacks, both procedure level and attack level cleanup (teardown) implementation is required.

If the emulator supports stopping an attack mid-runtime, then we expect the cleanup functionality to be invoked when the attack is interrupted. Cases in which cleanup of an interrupted attack is not possible are considered to be the result of a bug rather than an assessment criterion.






\subsubsection{Logs}
\emph{Threat emulator logs can help the operator understand whether the attack was executed successfully}, which stages of the attack were performed, whether the target-assets were acquired, etc. 
This criterion refers to the capability of a threat emulator to automatically produce and store log files. 
It can be of the following values:
\newline\noindent No -- There are no logging capabilities implemented. If needed, the operator can log the attack steps.
\newline\noindent Base -- Log entries are created at the beginning and end of the emulated attack execution.
\newline\noindent Adv. -- A log entry is generated for every executed procedure.


Note that if an emulator does not support multi-procedure attacks, the value for this criterion is set to [Base] by default (brackets are used to denote criteria values within the text).

\subsection{Operators}
\subsubsection{Required security expertise}

One of the goals of using threat emulators is to simplify the security assessment process. 
Ease of use and execution of attacks is an important factor when evaluating a threat emulator. 
Threat emulators that demand expert knowledge, complicated setups, or extensive configuration are counterproductive. 
We consider \emph{a good threat emulator as one that allows an operator without security expertise to execute and configure all built-in attacks.} Of course, the ability to add new procedures requires cyber security expertise. 
This criterion refers only to the expertise required to configure and execute attacks provided in the emulator's original package.
We define two levels of required security expertise as follows:
\newline\noindent Low -- Built-in attacks can be executed out of the box or by following step-by-step instructions provided with the emulator.
\newline\noindent High -- Provided instructions are not sufficient for an operator, who lacks cyber security knowledge, for configuring and executing attacks provided with the emulator.

Note that this criterion is different from the criterion of documentation described below, which considers the coverage and the level of detail of the documentation provided. 
Here we only consider the prior cyber security knowledge required to launch attacks.


\subsubsection{Documentation}
In addition to the need for an intuitive user interface, it is important to inspect the emulator's documentation. 
While some emulators consist of a poorly documented collection of scripts, some are well documented. 
Good documentation shortens the learning curve and facilitates full utilization of the emulator's capabilities. 
\emph{Even an operator that is unfamiliar with the emulator should be able to successfully create and execute attack scenarios and interpret the results relying only on his/her cyber security knowledge and the documentation provided.}

This criterion refers to the completeness of the documentation. It may be one of the following values:
\newline\noindent Full -- The documentation is sufficient for setting up all of the emulator's components, executing built-in attack scenarios, creating new ones, and interpreting the results.
\newline\noindent Miss -- Documentation of some of the critical functionality is missing.
\newline\noindent None -- The documentation is insufficient, making it difficult to set up the emulator or execute built-in attack scenarios.


\subsubsection{Interfaces}

It is important to map an emulator's interfaces in order to better suit the capabilities of the operators and organizational needs, such as the desired level of attack customization or the frequency of security assessment. 
Threat emulators can be operated through a simple command-line interface (CLI) or a graphical user interface (GUI). 
Novice operators may find a GUI more accessible and intuitive. Command-line tools are usually required for automating security assessment, but they have a steeper learning curve. 
Some emulators also support scripting (further denoted as SRP) to produce custom attack scenarios. 
A few emulators, such as Atomic Red Team and Metasploit, provide an application interface (API) to manage the attack scenarios. 
\emph{Ideally an emulator will provide all the above interfaces.}
The value of this criterion may consist of one or more of the following:
\newline\noindent GUI -- Some 
functionality is accessible through a GUI.
\newline\noindent CLI -- Some 
functionality is accessible through a CLI. 
\newline\noindent API -- Some 
functionality is accessible through an API.
\newline\noindent SRP -- The emulator supports scripting of attack scenarios or experiments. 


When all functionality relevant to the scenario's definition and execution is accessible through a GUI, CLI, or API, we denote this with an asterisk (*).

\section{Detailed Evaluation of Threat Emulators}
\label{sec:emulators}
In this section, we present a comprehensive evaluation methodology and review 11 open-source threat emulators based on the criteria described in Section~\ref{sec:criteria}.

\subsection{Evaluation Methodology}
\label{sec:methodology}
In this subsection, we describe the methodology for evaluating threat emulators with respect to the criteria presented in Section~\ref{sec:criteria}. 
The expert questionnaire for emulator assessment is described in Appendix~\ref{sec:questionnaire}. 
The step-by-step evaluation process is depicted in Fig.~\ref{fig:methodology} in Appendix~\ref{sec:evaluation-process-fig}.

First, we list the emulator components (such as the CNC server and agents) and \emph{prerequisites} (such as third party tools, libraries, and required privileges) needed to launch the threat emulator. 
If there is a need for third party tools or special privileges to be set up on the endpoints on which the emulated adversarial activity will be executed, then the values [3rd] or [Priv], respectively, will be selected for the \emph{prerequisites} criterion; otherwise, 
the value of this criterion is set to [$\emptyset$].

Next, we record whether or not the emulator uses agents to emulate adversarial activity. 
The \emph{OS compatibility} is determined by the emulator's agents if it uses them. 
In the absence of agents, the \emph{OS compatibility} is determined by the built-in procedures and third party tools required.
The CNC server, tools for analyzing the results, and other emulator components that can run on dedicated machines that are not a part of the original operational environment are not considered for \emph{OS compatibility}.

Then, we determine whether \emph{changes in the security array} are required to successfully set up the agents and install third party tools. For a threat emulator that works with agents we examine whether the agent can be installed without disabling the local firewall and antivirus software. 
At this stage we also check whether there are any antivirus alerts on third party tools or emulator files that should run on the endpoints.

After setting up the emulator, we examine the available \emph{interfaces} using which the emulated attacks can be defined and executed. 
We check whether the emulator has a GUI, API or CLI, and whether it supports scripting (SRP). 
For each \emph{interface}, we determine whether all functionality required for the attack scenario definition and execution is available via the \emph{interface}. 
If so, we add asterisk (*) to the \emph{interface}'s respective value.

To evaluate the threat emulator with respect to \emph{required security expertise}, we start by launching a built-in attack. 
If it is possible to launch the attack by using the documentation, without the need for arguments or configuration, this criterion will receive the value [Low]. 
Simple, meaningful names of procedures contribute to a [Low] \emph{required security expertise} criterion value. 
However, if the configuration of the attack (including choosing the relevant procedures) requires knowledge in cyber security beyond the documented configuration steps, this criterion will receive the value [High].

We also assess whether the emulator includes built-in \emph{multi-procedure attacks}. 
If it does, the \emph{multi-procedure attack} criterion receives the value [Built-in]. We execute any attack of the built-in multi-procedure attacks, and try to \emph{stop the attack mid-runtime}. 
We search for the stopping functionality in the available interfaces. 
For example, in cases in which the threat emulator has a GUI, we look for a ``stop'' button. 
If such functionality is provided and if, by activating it, all of the attack components are stopped, this criterion receives the value [Yes]; otherwise, it receives [No].

After the attack has stopped, we examine the \emph{logs} created, if such logs exist. 
If the \emph{log} describes each procedure that has been executed, this criterion will receive the value [Adv]. 
However, if only the attack metadata, such as start time, end time, endpoint, etc., was logged, without information about the execution of procedures, this criterion will receive the value [Base]. 
If none of the above exists, this criterion will receive the value [No].

After the attack has stopped, we also examine the emulator's \emph{cleanup} functionalities. 
If \emph{cleanup} is available, we verify that it is executed correctly, i.e., that the attack traces/artifacts are removed. 
For example, if a file was created, it should be deleted. 
If a registry key was modified, it should be restored to the previous value, etc. 
As noted earlier, send and forget actions cannot be \emph{cleaned up}. 
In addition, we examine the timing of the \emph{cleanup}. 
If the \emph{cleanup} is executed after each individual procedure, this criterion will receive the value [Proc]. 
If the \emph{cleanup} is executed after the execution of the entire attack scenario, this criterion will receive the value [Attk].

Next, we examine whether the emulator supports \emph{adding new procedures} and creating custom \emph{multi-procedure attacks}. 
First, we try \emph{adding a new procedure}, using the available interfaces. 
If the emulator is script-based, we add the new script to the relevant folder. 
We conclude that an emulator does not support \emph{adding new procedures} if it requires changing the emulator's source code; for example, inserting a new procedure into an existing script. 
To confirm whether the new procedure has been added successfully, we execute a simple attack that contains the procedure.

Then, we try creating a new \emph{multi-procedure attack}, either by using the available \emph{interfaces} or by combining available scripts. If a new procedure was added in the previous step, the new multi-procedure attack should include it. 
If the new custom multi-procedure attack was created and executed successfully, we assign the value [Custom] for the \emph{multi-procedure attack} criterion.

Using the new multi-procedure attack, or any built-in attack, we then examine the \emph{procedures' configuration} before execution. 
For each procedure that may require arguments, such as an endpoint's IP for lateral movement or data exfiltration, or a file name to search, we examine how these arguments can be specified. 
If the arguments are specified using a configuration file or the emulator's \emph{interfaces} before execution of the attack, the criterion value is set to [Med]. 
Otherwise, if the arguments are hardcoded, we change them in the code and set the criterion value to [Low]. 
Finally, we set the criterion to [High], if, when \emph{adding a new procedure}, the emulator enforces an API that supports future configuration changes.

To evaluate the \emph{TTP coverage} of an emulator, we are first required to list all of the built-in procedures the emulator provides.
We map each procedure to the techniques that it implements. 
Next, each technique is mapped to the tactics that can be achieved by using it. The mapping is performed according to the names of the procedures, techniques, and tactics using mainly the MITRE ATT\&CK Enterprise matrix. Note, that the names of the emulator's built-in procedures might not match the names on the MITRE ATT\&CK website. 
In such cases, we map the procedures to techniques according to their functionality. 
We summarize the mapping results in a vector of 10 dimensions, one for each tactic, except for the initial access and execution tactics. 
The value of each dimension is the total number of techniques that the threat emulator supports for the respective tactic. 
Initial access and execution are not included in our evaluation of threat emulators, because they are mainly covered by pen testing tools which are outside the scope of this survey.

Last but not least, we conclude with an assessment of the \emph{documentation} quality. 
By \emph{documentation} we refer to comments in a procedure's implementation, README files, text on the respective GitHub page or website, tooltips and help options in the emulator's interface, manuals, etc. 
Throughout the emulator evaluation process we were assisted by the available documentation. 
If every operation that we were required to perform was well documented, this criterion received the value [Full]. 
If one of the critical functions required for attack definition or execution was not well documented, this criterion received the value [Miss]. 
If the documentation does not include helpful instructions for setting up the emulator and executing built-in attack scenarios, we set the value of the \emph{documentation} criterion to [None].

\subsection{Highlights}
\label{sec:highlights}
Each threat emulator was manually examined to determine: 
(1) which of the MITRE ATT\&CK tactics and techniques are implemented, 
and (2) how each of the criteria discussed in Section~\ref{sec:criteria} are addressed. 
In this subsection we present the highlights for each threat emulator reviewed. A table containing a review of the emulators according to the discussed criteria is provided as supplementary material to the paper.
\footnote{\url{https://docs.google.com/spreadsheets/d/1zUq1QDHtZRxjde91S31\_InMK8alwf5\_xdruOR3qttQ8/}}


\subsubsection{Red Team Automation (RTA)}
Red Team Automation\footnote{https://github.com/endgameinc/RTA} is a framework comprised of Python scripts that blue teams can use to test their security mechanisms. 
It has no central components such as a CNC. 
Since RTA is written in Python, it can only be used on Python supported OSs and on endpoints that have Python installed on them. 
py2exe can be used to operate RTA on Windows OS without Python. 
To use all of the built-in functionality, RTA requires additional third party software. 
For example, in order to simulate lateral movement, psexec or xcopy are required.
RTA includes minimal documentation, with setup and execution instructions on its GitHub page\footnote{\url{https://github.com/endgameinc/RTA}} and in comments within its scripts. 
The scripts can be combined to create a multi-procedure attack. 
Additionally, it is possible to edit procedures or add new ones using Python. Hence, to be used effectively, RTA requires Python knowledge.
The emulated attack execution and the changes made to the endpoints are displayed onscreen, but there is no log file. 
RTA has a cleanup process which is useful for reverting the changes made in each procedure.

RTA scripts implement a reasonable number of techniques from MITRE's ATT\&CK matrix. 
However, the coverage of the tactics is not uniform. 
For example, persistence and defense evasion tactics are well covered, with 10 and 20 techniques respectively, while lateral movement has only three techniques implemented in RTA. 
RTA can also generate network traffic that mimics communication with the attacker's CNC.
RTA provides both procedures that antiviruses usually do not detect and common procedures that can be detected by antiviruses. 
RTA does not require disabling security tools in order to launch most of the procedures.
Overall, RTA is suitable for testing an entire network and is capable of simulating real-life threats, due to the diversity of the tactics and techniques. 
The fact that this tool is Python-based makes it easier to add advanced techniques and more sophisticated attacks.

\subsubsection{Metasploit}
Metasploit\footnote{\url{https://www.metasploit.com/}} is one of the most commonly used pen testing toolkits. 
It focuses on initial access and execution tactics and contains a large library of vulnerabilities and exploits~\cite{holik2014effective}.
Metasploit can also be used as a threat emulator for security assessment thanks to the post-compromise tools it contains. 
Metasploit attack modules may deliver a payload to the compromised endpoint in order to perform post-compromise activities. 
Meterpreter is Metasploit's most well-known payload. 
Meterpreter provides good coverage of post-compromise tactics, including discovery, collection, persistence, privilege escalation, and more. 
It employs third party tools, such as psexec and BruteForce, for lateral movement.
Metasploit does not have a library of built-in multi-procedure attacks, but extensive attack scenarios that mimic real APTs can be scripted. 
For example, AutoTTP\footnote{\url{https://github.com/jymcheong/AutoTTP}} is a scripting package that enables generating and executing multi-procedure attack scenarios using the APIs of Metasploit and Empire\footnote{Empire is a PowerShell and Python post-compromise agent \url{https://github.com/EmpireProject/Empire}}.  
AutoTTP organizes the attack procedures along an attack lifecycle suggested by its developer Jym Cheong.
Most of the attacks performed by Metasploit are not detected by standard antivirus tools.

Metasploit server components are part of the Kali Linux OS, but they can also be installed on the Windows OS.
Metasploit target endpoints may run the Windows OS, the Linux OS, and MacOS. 
An operator needs to have extensive cyber security knowledge and some knowledge of Ruby in order to both launch built-in attacks and add new attack procedures. 
To make it easier for the operator, a GUI version of Metasploit, called Armitage, has been developed\footnote{http://www.fastandeasyhacking.com/}.



\subsubsection{Invoke-Adversary}
Invoke-Adversary\footnote{https://github.com/CyberMonitor/Invoke-Adversary} is a PowerShell script used by blue teams to test their security mechanisms. Since Invoke-Adversary is written in PowerShell, it can be used only on PowerShell supported OSs and on endpoints that have PowerShell installed and enabled on them.

Invoke-Adversary's main menu is based on MITRE's ATT\&CK tactics list. For each attack, the menu allows the operator to choose a single technique for each tactic. 
Invoke-Adversary implements six to 10 procedures for each of the following tactics: persistence, discovery, credential access, defense evasion, command and control, and execution. 
However, Invoke-Adversary provides only one procedure for the collection tactic. 
Lateral movement is not supported.
Three of the six procedures provided by Invoke-Adversary for the credential access tactic are detected by an antivirus. 
The other three (LSASS dump-based) procedures are rarely detected by antiviruses. 
Other procedures provided are common and can be detected by antiviruses.

Because it is a PowerShell script, Invoke-Adversary can be modified by an operator that has sufficient knowledge in PowerShell and cyber security. 
Such an operator can edit procedures, add new custom procedures, and compose multi-procedure attacks.
Invoke-Adversary does not provide cleanup functionality.
It outputs a log to the console at the end of attack execution, including procedures' results (if successful) and the reason for failures (if unsuccessful). 

Overall, Invoke-Adversary is suitable for testing a standalone endpoint, but operators are required to have cyber security expertise in order to execute attacks that combine more than one procedure.

\subsubsection{Blue Team Training Toolkit (BT3)}
BT3\footnote{\url{https://www.bt3.no/}} is a platform that generates network traffic that mimics malicious activity. 
It is primarily used as blue team training kit. 
BT3 offers free components and components that must be paid for. 
The free components include 14 network attack simulations and 10 methods for generating dummy malicious files using hash collision. 
Some BT3 components require additional third party tools such as LHOST, SSL, etc.
The attack components cannot be combined into a multi-procedure attack scenario.

BT3 must be installed on a Linux server, however the network simulation can target endpoints with any OS. 
Hash collision can only be performed on the Linux server.
BT3 does not require Linux knowledge. It provides install.sh script for easy setup. 
However, in order to execute a malicious network traffic simulation, the operator should be able to recognize attack procedures and the required configurations based on the procedures' names. 
Consequently, cyber security knowledge is required to correctly operate the network simulations. 
The BT3 website provides sufficient documentation on all BT3 components. 
BT3 does not have a cleanup functionality.
Files created from hash collision need to be manually removed by the operator. 


Overall, BT3 is suitable for testing an IDS or hash-based antivirus. 
The fact that BT3 does not contain multi-procedure attacks and does not support various tactics and techniques makes it difficult to use for threat emulation.

\subsubsection{Advanced Persistent Threat (APT) Simulator}
APTSimulator\footnote{\url{https://github.com/NextronSystems/APTSimulator}} is a Windows batch script that aims to be as simple as possible with little user interaction. 
APTSimulator runs on Windows with no additional setup requirements, 
but it doesn't support other OSs.

The procedures' documentation, including execution instructions, appears in comments inside the code. 
However, with sufficient expertise, it is possible to add new procedures and create new attacks by editing and combining batch scripts. 
Sophisticated procedures and techniques can be created by adding third party tools to APTSimulator's folder and writing an appropriate batch script that utilizes them.
Most of the procedures implemented in APTSimulator were detected by an antivirus. 
The procedures detected by an antivirus are listed on the tool's GitHub page. 
Currently, APTSimulator implements techniques for persistence and discovery tactics (eight and seven techniques, respectively), 
but it lacks lateral movement capabilities and CNC component. 
Every procedure executed provides on-screen logging, but there is no log file or cleanup.

Overall, APTSimulator is suitable for endpoint 
testing. 
It can be deployed and ran in minutes on Windows environment and it provides a basic testing for the security of an endpoint.

\subsubsection{CALDERA}
CALDERA\footnote{https://github.com/mitre/caldera} is a threat emulator developed by the MITRE organization. 
It is designed to test the security of a Windows system~\cite{Applebaum2016}.
CALDERA includes a remote access agent, a database, and server components. 
CALDERA's server can be installed on a Windows server or Windows 10 platforms. 
CALDERA's agents can run on Windows. 
The installation and configuration process of CALDERA’s components takes longer than that of most of the other threat emulators reviewed. 
In order to use CALDERA's remote agent, the antivirus software running on the endpoint must be disabled.

Once installed, CALDERA's graphical interface allows the operator to control all of the agents and obtain visual feedback about an ongoing attack. 
All of the attacks in CALDERA are taken from the MITRE ATT\&CK matrix, and information about the procedures can be found in the MITRE documentation. Attack procedures can be chosen from the CALDERA library and combined together to create custom multi-procedure attacks. 
CALDERA gives the operator the option to import his/her own RAT to perform additional procedures or have different CNC communication. 
CALDERA's lateral movement capabilities allow choosing the first compromised machine (``patient zero'') out of the endpoints with CALDERA's agents. 
Other tactics that CALDERA covers are persistence, privilege escalation, and discovery with 7, 6, and 11 techniques respectively. 
In addition, CALDERA provides a built-in cleanup functionality.

Overall, CALDERA provides a user friendly interface and functionality to compile new attacks from existing procedures.
CALDERA is most suitable for those who search for a threat emulator with a large variety of built-in features, lateral movement, and CNC communication using a RAT, and a large set of discovery procedures.

\subsubsection{Infection Monkey}
Infection Monkey\footnote{\url{https://github.com/guardicore/monkey}} is a threat emulator used for testing and assessing the defenses against lateral movement and discovery tactics  
(mostly initial access).
Infection Monkey has a CNC component, and a RAT must be installed on the endpoints. 
Since antivirus tools consider the RAT used by Infection Monkey as a threat, antivirus tools must be disabled for Infection Monkey to function properly.
Infection Monkey's server can be installed on any OS, and no additional knowledge is required to launch an attack; its RAT component can be installed on Windows and Linux OSs (but not on MacOS).

The Infection Monkey RAT can be used to choose the first compromised machine from the endpoints. 
Infection Monkey reports about machines that were infected in the process of lateral movement. 
Infection Monkey was designed with the aim of testing defenses against lateral movement, but it lacks a diverse range of tactics and supports few procedures.
The techniques and exploits used by Infection Monkey come from real-world scenarios, so it closely simulates a real attacker's behavior when it comes to lateral movement. 
Notably, most of the procedures, if executed separately without the RAT, would have remained undetected by antiviruses.
The progress of each executed attack is logged in the interface, and a detailed review of the results is provided. 
Infection Monkey provides a cleanup option which is useful when examining live systems.

Overall, Infection Monkey is a good threat emulator for testing network defenses. 
Moreover, its easy installation process together with its graphical user interface, cleanup, logging, and OS compatibility make it a suitable choice for novice operators that are looking for an easy way to evaluate the network environment.

\subsubsection{PowerSploit}
PowerSploit\footnote{\url{https://github.com/PowerShellMafia/PowerSploit}} is a collection of PowerShell-based scripts aimed at various security assessment tasks. 
Since PowerSploit is based on PowerShell, it runs natively on any Windows OS with PowerShell installed. 
In addition, there are add-ons that can be installed on MacOS and Linux OS to enable using PowerShell.

PowerSploit does not support adding new procedures, but PowerSploit's scripts can be combined into a custom multi-procedure attack. 
PowerSploit supports a high number of sophisticated procedures. 
As a result it requires cyber security expertise in addition to PowerShell knowledge. 
PowerSploit supports persistence, privilege escalation, defense evasion, and discovery tactics with 18, 9, 18, and 18 techniques respectively. 
It also supports credential access, collection, and lateral movement tactics with seven, seven, and six techniques, respectively.
Every executed procedure provides on-screen logging, but there is no log file. 
PowerSploit lacks cleanup and logging functionalities.

Overall, PowerSploit can be a quick and easy baseline assessment tool. It is well documented, and it contains a large number of procedures, and more specifically, many procedures that try to avoid getting caught by antiviruses.

\subsubsection{DumpsterFire}
DumpsterFire\footnote{\url{https://github.com/TryCatchHCF/DumpsterFire}} is a platform for building attacks using techniques that are largely not part of the MITRE ATT\&CK matrix. 
For example, the creation of 5,000 files, searching for hacking tools in Google, and running a video in a loop on YouTube. 
DumpsterFire can only be executed on Linux systems. 
DumpsterFire does not have a cleanup option, and documentation only appears on their GitHub page and within the procedure scripts in the form of comments. 
Logs of the execution of each procedure are shown on the CLI.

A multi-procedure attack is built by choosing techniques from a menu. DumpsterFire provides a set of built-in multi-procedure attacks packed with a well-built CLI. 
In addition, DumpsterFire implements a variety of discovery and credential access techniques (13 and 6 techniques, respectively). However, DumpsterFire lacks CNC and lateral movement techniques. Overall, DumpsterFire supports a small number of techniques covering just a few tactics. 
However, the techniques it contains and its simple interface make DumpsterFire useful for examining the basic security of Linux endpoints in an easy and accessible manner.

\subsubsection{Uber's Metta}
Metta\footnote{\url{https://github.com/uber-common/metta}} is a threat emulator developed by Uber Technologies. 
Its main purpose is assessing endpoint security, but it also includes a set of network security testing procedures. 
Uber's Metta can be executed on endpoints with the Linux OS, the Windows OS, and MacOS. 
In order to use Metta, an organization needs to install a Redis server, Python 2.7, and Vagrant, which makes the installation process a bit more complicated than that of other threat emulators.

Metta provides built-in attacks, each of which executes all of the techniques that achieve a specific tactic. 
For example, an attack that consists of all techniques that collect a user's data. 
Such an attack does not emulate the complete attack lifecycle, but rather focuses on a wide assessment of specific defense targets.
Metta also provides the operator with the ability to create custom multi-procedure attacks. 
However, the operator should have coding experience in order to add, launch, and manage attacks.
Metta supports CNC and lateral movement tactics. 
But lateral movement procedures are only implemented for Linux OS. 
Metta also provides various techniques for achieving discovery, credential access, and defense evasion tactics (10, 7, and 7 techniques, respectively). 
Persistence, privilege escalation, collection, and exfiltration tactics are supported by Metta with just a few techniques. 
Most of the procedures implemented in Metta were not detected by antivirus tools.
For each attack there is on-screen logging, and a log file is generated at the end of the attack. 
Metta has no cleanup functionality.

Overall, Metta is suitable for operators who 
need to test specific parts of endpoint security across many platforms.

\subsubsection{Atomic Red Team}
Atomic Red Team is a threat emulation library of lightweight security tests that can be rapidly executed by security teams\footnote{\url{https://atomicredteam.io/},\url{https://github.com/redcanaryco/atomic-red-team}}. 
Attack scenarios can be executed using command line, PowerShell, and Shell.
It is compatible with all major operating systems.
Atomic Red Team also provides an API written in Ruby.

Atomic Red Team supports multiple techniques for achieving persistence, privilege escalation, defense evasion, credential access, and discovery tactics (20, 9, 24, 9, and 16 techniques, respectively). 
It also includes procedures for lateral movement and simulates CNC communication. 
All procedures are mapped to the MITRE ATT\&CK matrix. 
It is possible to add new custom procedure scripts. 
In addition, Atomic Red Team uses third party tools, such as mimikatz, for performing credential dumping techniques. 
Most of the procedures are undetected by antivirus tools. 
Atomic Red Team has cleanup at the procedure level. 
It also provides on-screen logging which outputs information about the currently running procedure during attack execution. 
Documentation is provided only as comments in the procedures' scripts.

One of the main objectives of Atomic Red Team is allowing the operator to execute procedures quickly, with minimal installation, however Atomic Red Team may require cyber security knowledge in order to be used effectively.


\section{Taxonomy of Threat Emulators}
\label{sec:taxonomies}

In this section, we summarize the findings from our review of the emulators in a taxonomy that highlights their qualities. 
The emulator review questionnaire (see Appendix~\ref{sec:questionnaire}) was completed by a review panel, based on factual assessment of the emulators. 
The criteria values were inferred from the questionnaire (see supplementary material) and aggregated according to the criteria hierarchy (Fig.~\ref{fig:criteria}).

\begin{figure}
\includegraphics[width=\columnwidth]{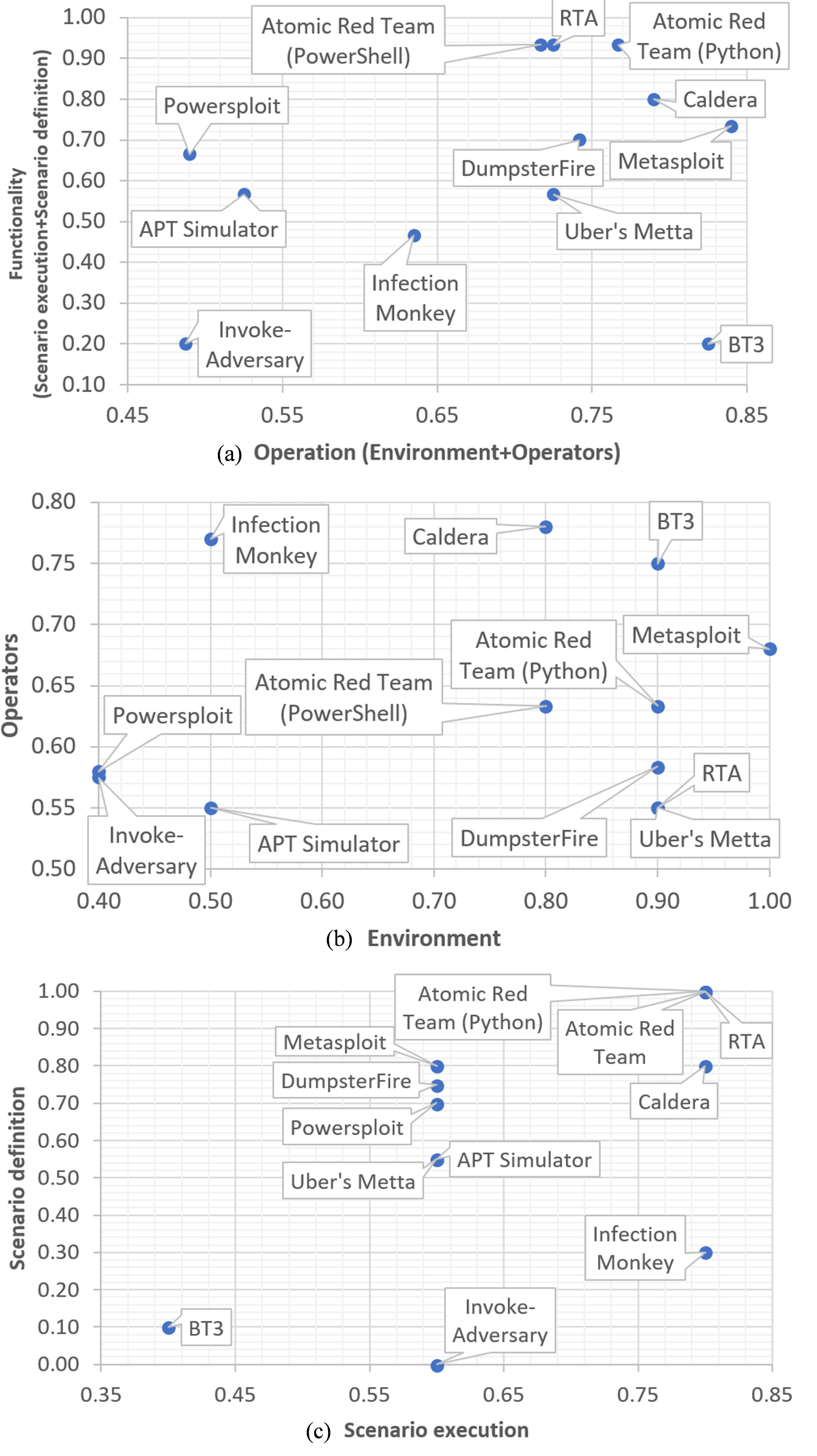}
\caption{\label{fig:taxonomy}Qualities of the threat emulators} 
\end{figure}

Fig.~\ref{fig:taxonomy} presents the strengths and weaknesses of the emulators with respect to four criteria dimensions: operators, environment, scenario definition, and scenario execution. 
Inspecting the highest aggregation level (functionality vs. operation) in Fig.~\ref{fig:taxonomy}a, we can identify four outstanding threat emulators in the top right corner: RTA, Atomic Red Team, CALDERA, and Metasploit, which form the Pareto frontier. 
These emulators exhibit the best trade-off between the ease of setting up and operating the emulator in a wide range of environments (operation) and the ability to define and execute diverse attack scenarios (functionality).

Further inspecting the operator dimension using Fig.~\ref{fig:taxonomy}b, we can identify Infection Monkey, CALDERA, and BT3 as the emulators that are most operator friendly. 
Metasploit is outstanding in terms of environment criteria, supporting all operating systems with minimal prerequisites.

Atomic Red Team, RTA, CALDERA, and Infection Monkey provide the most features related to controlling and logging attack execution (Fig.~\ref{fig:taxonomy}c). 

With respect to the diversity and flexibility of attack scenario definition, it is also important to consider the summary of tactics as presented in Table~\ref{fig:tactics}. 
Higher diversity means that the red team is able to create more attack scenarios, and the blue team may gain better understanding of the security array's limitations. 
The top emulators according to the scenario definition criteria also have the highest diversity of supported techniques (except for DumpsterFire).

It is apparent from Table~\ref{fig:tactics} that Metasploit and RTA pay much less attention to the discovery tactic than their competitors.
On the other hand CALDERA, Uber's Metta, and DumpsterFire emulators implement many discovery techniques (which are usually noisy), while providing less support for persistence and defense evasion. 
These three emulators may be a good choice when evaluating anomaly detection methods.

\begin{table}
\centering
\caption{Summary of tactics}
\label{fig:tactics}
\includegraphics[width=0.9\columnwidth]{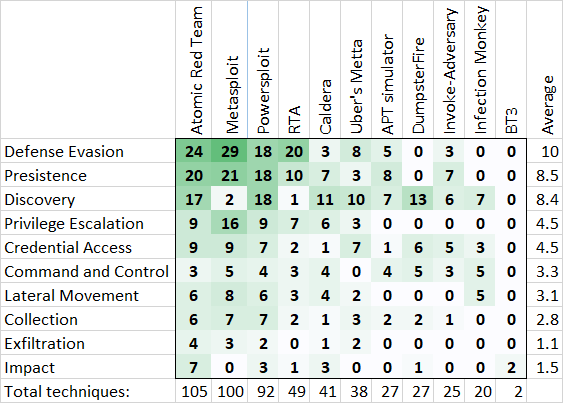}
\end{table}



\subsection{Selecting a Threat Emulator}


Based on the taxonomy above and the threat emulation use cases, we formulate guidelines for choosing the appropriate threat emulator.

\noindent\emph{Training.}
Since training often takes place in a controlled environment, like a cyber range, the environmental requirements are looser than, for example, in the organizational security assessment use case. 
Because emulators are used to train blue teams in handling a variety of security events, they should be scored high in the attack scenario definition criteria. 
As a case in point, emulators that support a variety of tactics and techniques are preferred. 
When training exercises are managed by experienced instructors, the operators criteria do not need to be considered. 
As a result, Metasploit, RTA, Atomic Red Team, and CALDERA are the top candidates for training exercises (see Fig.~\ref{fig:taxonomy}c). 
An experienced red team may prefer DumpsterFire over PowerSploit despite the lower number of tactics, because DumpsterFire supports better attack customization. 
However, in the case of a blue team self-training, a qualified red team may be absent; in that scenario, an operator without extensive security expertise will prefer emulators scoring high in both the scenario definition and operators criteria, such as CALDERA (see Fig.~\ref{fig:taxonomy}). 

\noindent\emph{Security Tool Assessment.}
Emulators scoring high on the environment criteria widen the range of tools that can be assessed using the emulator, and ease of operation increases the productivity of the assessment. 
Attack definition criteria and in particular the variety of supported tactics and techniques are highly important in challenging the security tool in as many scenarios as possible. 
Configurability is also important for supporting the repeated execution of simulated attacks with different security products to compare and contrast them. 
Although emulators' advanced logging capabilities are beneficial during security tool assessment, stopping the attack mid-runtime and cleanup are not required, making the execution criteria less important. 
Of Atomic Red Team, RTA, CALDERA, Metasploit, DumpsterFire and PowerSploit which all score high on the scenario definition criteria, Metasploit and CALDERA exhibit the best operator vs. environment trade-off.

\noindent\emph{Organizational Security Assessment}
implies the execution of a threat emulator within the operational organizational environment. 
In such settings, all scenario execution criteria, especially cleanup and the ability to stop an emulated attack mid-runtime, are crucial for troubleshooting the assessment process. 
All environment criteria (OS compatibility, special privileges, etc.) are important for seamless integration in the existing organizational environment and reduce the disruption of the organization's operations. Based on the scenario execution and environment criteria, Atomic Red Team and RTA score the highest, 
because they support cleanup and have loose environment constraints (see Figs.~\ref{fig:taxonomy}c and~\ref{fig:taxonomy}b).

\noindent\emph{What-If Analysis}
requires repeated, preferably automated, execution of an attack scenario with slight changes in either the security array or attack definition. One of the main objectives of what-if analysis is assessment of the potential impact of a security event on the organization. 
As such it is best performed on the organizational premises, resulting in threat emulators that intertwine with the operational organizational environment. 
Overall, the most important criteria for what-if analysis are the environment criteria, as well as the cleanup and procedures' configuration criteria. 
Atomic Red Team and RTA support cleanup and also have the highest score with respect to the procedures' configuration (see criteria values in the supplementary material) and are second only to Metasploit with respect to the environment criteria (see Fig.~\ref{fig:taxonomy}b).

\noindent\emph{General Considerations.}
As can be seen in Fig.~\ref{fig:taxonomy}b, a novice red team member who is technology savvy but doesn't have cyber security expertise would prefer the Infection Monkey or CALDERA threat emulator.

If the purpose is to determine whether a blue team can handle advanced persistence threats (APT), the threat emulator should support a high level of attack scenario customization, including adding new procedures and custom multi-procedure attacks. 
APT emulators should have high TTP coverage in general and in particular support many lateral movement and defense evasion techniques; as can be seen in Fig.~\ref{fig:taxonomy}c, the most appropriate emulators are Metasploit, Atomic Red Team, RTA, and CALDERA. 
When the focus is testing advanced adversarial techniques, such as process injection, file encryption, or credential dumps, PowerSploit is a better choice than CALDERA.






\section{Summary}
\label{sec:summary}
Threat emulators are valuable tools for training security personnel, security assessment, and what-if analysis. 
In this article we reviewed 11 open-source threat emulators and presented a detailed and comprehensive evaluation methodology for qualitatively and quantitatively comparing them.

The review is based on well-defined criteria organized in four dimensions as discussed in Section~\ref{sec:criteria}. 
The results identify four open-source threat emulators (Atomic Red Team, RTA, CALDERA and Metasploit) that lead the way according to our summary of the criteria. 
Also worth mentioning is Infection Monkey, which stands out as an operator friendly threat emulator. 
Based on the taxonomy presented, we formulate guidelines for selecting the most appropriate threat emulator in a given use case. 
Finally, our evaluation methodology and taxonomy highlight the need to address the following aspects which are crucial for the further development of threat emulators:
\begin{itemize}
    \item Cleanup and configurability are important in order to repeat and automate the execution of attack scenarios during security tool assessment and what-if analysis. 
    \item An emulator should support cleanup after the completion of the attack scenario, like CALDERA, Atomic Red Team, and Infection Monkey do, rather than after each individual procedure.
    \item An API, currently provided by Atomic Red Team, CALDERA, and Metasploit, facilitates integration between the threat emulators and organizational security array, thus enabling periodic and systematic security assessment.
    
    \item It is important to provide a GUI and ready to execute multi-procedure attacks for novice operators as well as a CLI to support automation and advanced customization capabilities.
\end{itemize}

\bibliographystyle{plain}
\bibliography{references}

\section{APPENDIX: Emulator Evaluation Questionnaire}
\label{sec:questionnaire}
The following is a conditional questionnaire where the answers to certain questions may affect the relevance of other questions. 
The term \emph{attack} may refer to single or multi-procedure attack scenarios. Explicit reference to \emph{multi-procedure attacks} or to \emph{procedures} is provided where relevant.

\subsection{Environment}
Questions in this section relate to the interactions between the emulator and the environment in which it is deployed.
The following questions relate to the emulator's main components and dependencies. 

\noindent\begin{tabular}{|p{6.5cm}|c|c|}
\hline
 & \textbf{Yes} & \textbf{No} \\
\hline
\question{\label{q:agent}Does the emulator use agents?} & $\Box$ & $\Box$\\
\hline
\question{\label{q:script}Does the emulator execute scripts on the endpoints?} & $\Box$ & $\Box$\\
\hline
\question{\label{q:3rd}Does the emulator execute third party tools on the endpoints?} & $\Box$ & $\Box$\\
\hline
\end{tabular}

\subsubsection{OS compatibility}
W, L, and M refer to the Windows, Linux, and Mac operating systems respectively.

\noindent
\begin{tabular}{|p{6cm}|c|c|c|}
\hline
 & \textbf{W} & \textbf{L} & \textbf{M} \\
\hline
\question{[Q\ref{q:agent}=Yes] Are the emulator's agents compatible with the following operating systems?} & $\Box$ & $\Box$ & $\Box$ \\
\hline
\question{[Q\ref{q:script}=Yes] Does the emulator contain procedures compatible with the following operating systems?} & $\Box$ & $\Box$ & $\Box$ \\
\hline
\question{[Q\ref{q:3rd}=Yes] Does the emulator use third party tools compatible with the following operating systems?} & $\Box$ & $\Box$ & $\Box$ \\
\hline
\end{tabular}
 
\subsubsection{Changes to the security array}~~~
The following questions refer to built-in attacks only. 

\noindent
\begin{tabular}{|p{6.5cm}|c|c|}
\hline
 & \textbf{Yes} & \textbf{No} \\
\hline

\question{\label{q:FW} Did the system's firewall interrupt the emulator's workflow?}
& $\Box$ & $\Box$ \\
\hline
\question{[Q\ref{q:FW}=Yes] Did the system's firewall block the connection between the CNC and the RAT?} & $\Box$ & $\Box$ \\
\hline
\question{\label{q:AV} Did the system's real-time antivirus interrupt with the emulator's workflow?}
& $\Box$ & $\Box$ \\
\hline
\question{[Q\ref{q:AV}=Yes] Did the system's real-time antivirus interrupt any RAT functionality?}  & $\Box$ & $\Box$\\
\hline
\question{[Q\ref{q:AV}=Yes][Q\ref{q:agent}=Yes] Did the system's real-time antivirus interrupt the agents' functionality?}  & $\Box$ & $\Box$\\
\hline
\question{[Q\ref{q:AV}=Yes][Q\ref{q:script}=Yes] Did the system's real-time antivirus interrupt the scripts' functionality?} (i.e., delete PowerShell scripts) & $\Box$ & $\Box$\\
\hline
\question{[Q\ref{q:AV}=Yes][Q\ref{q:3rd}=Yes] Did the system's real-time antivirus interrupt the third party tools' functionality?} & $\Box$ & $\Box$ \\
\hline
\end{tabular}

\subsubsection{Prerequisites}~~~
The following questions refer to built-in attacks only. 

\noindent
\begin{tabular}{|p{6.5cm}|c|c|}
\hline
  & \textbf{Yes} & \textbf{No} \\
\hline
\question{[Q\ref{q:agent}=Yes] Do the emulator's agents require special privileges from the systems to take advantage of their full functionality?} & $\Box$ & $\Box$\\
\hline
\question{[Q\ref{q:script}=Yes] Do the emulator's scripts require special privileges from the systems to take advantage of their full functionality?} & $\Box$ & $\Box$\\
\hline
\question{[Q\ref{q:3rd}=Yes] Do the third party tools require special privileges from the systems to take advantage of their full functionality?} & $\Box$ & $\Box$\\
\hline
\end{tabular}

\subsection{Operators}
Questions in this section assess the ease of operating the emulator. 

\subsubsection{Required security expertise}
Consider a novice red team member, such as a software engineer or computer science student, who is technology savvy but doesn't have cyber security expertise. 

\noindent\begin{tabular}{|p{6.5cm}|c|c|}
\hline
 & \textbf{Yes} & \textbf{No} \\
\hline
\question{Can the novice red team member execute built-in attacks by following the instructions provided with the emulator?}
 & $\Box$ & $\Box$ \\
\hline
\end{tabular}

\subsubsection{Documentation}~~~

\noindent
\begin{tabular}{|p{6.5cm}|c|c|}
\hline
  & \textbf{Yes} & \textbf{No} \\
\hline
\question{\label{q:doc}Does the emulator have any kind of documentation?} & $\Box$ & $\Box$\\
\hline
\question{[Q\ref{q:doc}=Yes] Is the documentation sufficient for setting up all of the emulator's components?} & $\Box$ & $\Box$ \\
\hline
\question{[Q\ref{q:doc}=Yes] Is the documentation sufficient for launching built-in attacks?} & $\Box$ & $\Box$ \\
\hline
\question{[Q\ref{q:doc}=Yes][Q\ref{q:newproc}=Yes] Is the documentation sufficient for creating new custom procedures?} & $\Box$ & $\Box$ \\
\hline
\question{[Q\ref{q:doc}=Yes][Q\ref{q:multiprocnew}=Yes] Is the documentation sufficient for creating new multi-procedure attack scenarios?} & $\Box$ & $\Box$ \\
\hline
\question{[Q\ref{q:doc}=Yes] Does the documentation describe how to interpret attack execution results?} & $\Box$ & $\Box$ \\
\hline
\end{tabular}

\subsubsection{Interface}
In the following matrix indicate whether or not the functionality (row) is available through the UI (column). 

\noindent
\begin{tabular}{|p{5.1cm}|c|c|c|}
\hline
  & \textbf{GUI} & \textbf{CLI} & \textbf{API}\\
\hline
\question{Executing attacks} & $\Box$ & $\Box$& $\Box$\\
\hline
\question{\label{q:confui}Configuring procedures} & $\Box$ & $\Box$& $\Box$\\
\hline
\question{Stopping attacks mid-runtime} & $\Box$ & $\Box$& $\Box$\\
\hline
\question{[Q\ref{q:log}=Yes] Accessing logs} & $\Box$ & $\Box$& $\Box$\\
\hline
\question{[Q\ref{q:newproc}=Yes$\vee$\ref{q:procsrp}=Yes] Adding new custom procedures} & $\Box$ & $\Box$& $\Box$\\
\hline
\question{[Q\ref{q:multiprocnew}=Yes] Adding new multi-procedure attacks} & $\Box$ & $\Box$& $\Box$\\
\hline
\end{tabular}


\noindent
\begin{tabular}{|p{6.5cm}|c|c|}
\hline
  & \textbf{Yes} & \textbf{No}\\
\hline
\question{Does the emulator support attack scripting?} & $\Box$ & $\Box$\\
\hline
\end{tabular}

\subsection{Scenario execution}
Questions in this section assess the emulator's functionality related to attack scenario execution.



\subsubsection{Cleanup}~

\noindent
\begin{tabular}{|p{6.5cm}|c|c|}
\hline
\textbf{}  & \textbf{Yes} & \textbf{No} \\
\hline
\question{\label{q:clean}Does the emulator have cleanup functionality?} &  $\Box$  &  $\Box$\\
\hline
\question{[Q\ref{q:clean}=Yes] Does cleanup occur immediately after the relevant attack procedure?} & $\Box$ & $\Box$ \\
\hline
\question{[Q\ref{q:clean}=Yes][Q\ref{q:multiprocbuiltin}=Yes$\vee$Q\ref{q:multiprocnew}=Yes] Can cleanup of the relevant procedures be executed only at the end of a multi-procedure attack?} & $\Box$ & $\Box$ \\
\hline
\end{tabular}

\subsubsection{Logs}~~~

\noindent
\begin{tabular}{|p{6.5cm}|c|c|}
\hline
\textbf{}  & \textbf{Yes} & \textbf{No} \\
\hline
\question{\label{q:log}Does the emulator have logging capabilities?}   & $\Box$  &  $\Box$\\
\hline
\question{[Q\ref{q:log}=Yes] Is every executed procedure logged during an attack?} & $\Box$ & $\Box$ \\
\hline
\end{tabular}

\subsection{Scenario definition}
Questions in this section assess the functionality related to defining and configuring attack scenarios.

\subsubsection{Adding new procedures}~

\noindent
\begin{tabular}{|p{6.5cm}|c|c|}
\hline
\textbf{}  & \textbf{Yes} & \textbf{No} \\
\hline
\question{\label{q:newproc}Does the emulator support adding new custom procedures through any of the interfaces?} &  $\Box$ & $\Box$\\
\hline
\question{\label{q:procsrp}[Q\ref{q:newproc}=No] Are procedures implemented using scripts?} & $\Box$ & $\Box$ \\
\hline
\question{\label{q:srpcoll}[Q\ref{q:newproc}=No][Q\ref{q:procsrp}=Yes] Can a new script be added to the collection of procedures?} & $\Box$ & $\Box$ \\
\hline
\end{tabular}

\subsubsection{Procedures' configuration}~

\noindent
\begin{tabular}{|p{6.5cm}|c|c|}
\hline
\textbf{}  & \textbf{Yes} & \textbf{No} \\
\hline
\question{\label{q:conff}Can the emulator procedures be configured through configuration files?} &  $\Box$ &  $\Box$ \\
\hline
\question{\label{q:repeat}[Q\ref{q:conff}=No] Does the emulator support repeated execution of the same attack with different parameters?} &  $\Box$ &  $\Box$ \\
\hline
\question{[Q\ref{q:newproc}=Yes$\vee$Q\ref{q:srpcoll}=Yes] Can new custom procedures be configured using the same methods as built-in procedures (e.g., various interfaces or configuration files)?} & $\Box$ & $\Box$ \\
\hline
\end{tabular}

\subsubsection{Multi-procedure attacks}~

\noindent
\begin{tabular}{|p{6.5cm}|c|c|}
\hline
\textbf{}  & \textbf{Yes} & \textbf{No} \\
\hline
\question{\label{q:multiprocbuiltin}Does the emulator include built-in multi-procedure attacks?} & $\Box$ & $\Box$ \\
\hline
\question{\label{q:multiprocnew}Does the emulator support adding new custom multi-procedure attacks?} & $\Box$ & $\Box$ \\
\hline
\end{tabular}

\subsubsection{TTP coverage}
In order to answer the following questions, fill-in the MITRE ATT\&CK matrix\footnote{\url{https://attack.mitre.org/matrices/enterprise/}} for the emulator considering only built-in procedures.   
\noindent
\begin{tabular}{|p{4.5cm}|c|c|c|}
\hline
\textbf{}  & \textbf{$<$6} & \textbf{6 to 8} & \textbf{$>$8}\\
\hline
\question{The number of tactics is} & $\Box$ & $\Box$ & $\Box$\\
\hline
\textbf{}  & \textbf{$<$20} & \textbf{20 to 40} & \textbf{$>$40}\\
\hline
\question{The number of techniques is }& $\Box$ & $\Box$ & $\Box$\\
\hline
\end{tabular}

\section{APPENDIX: Threat Emulators Evaluation Process}
\label{sec:evaluation-process-fig}
\begin{figure*}
    \centering
    \includegraphics[width=1.0\textwidth]{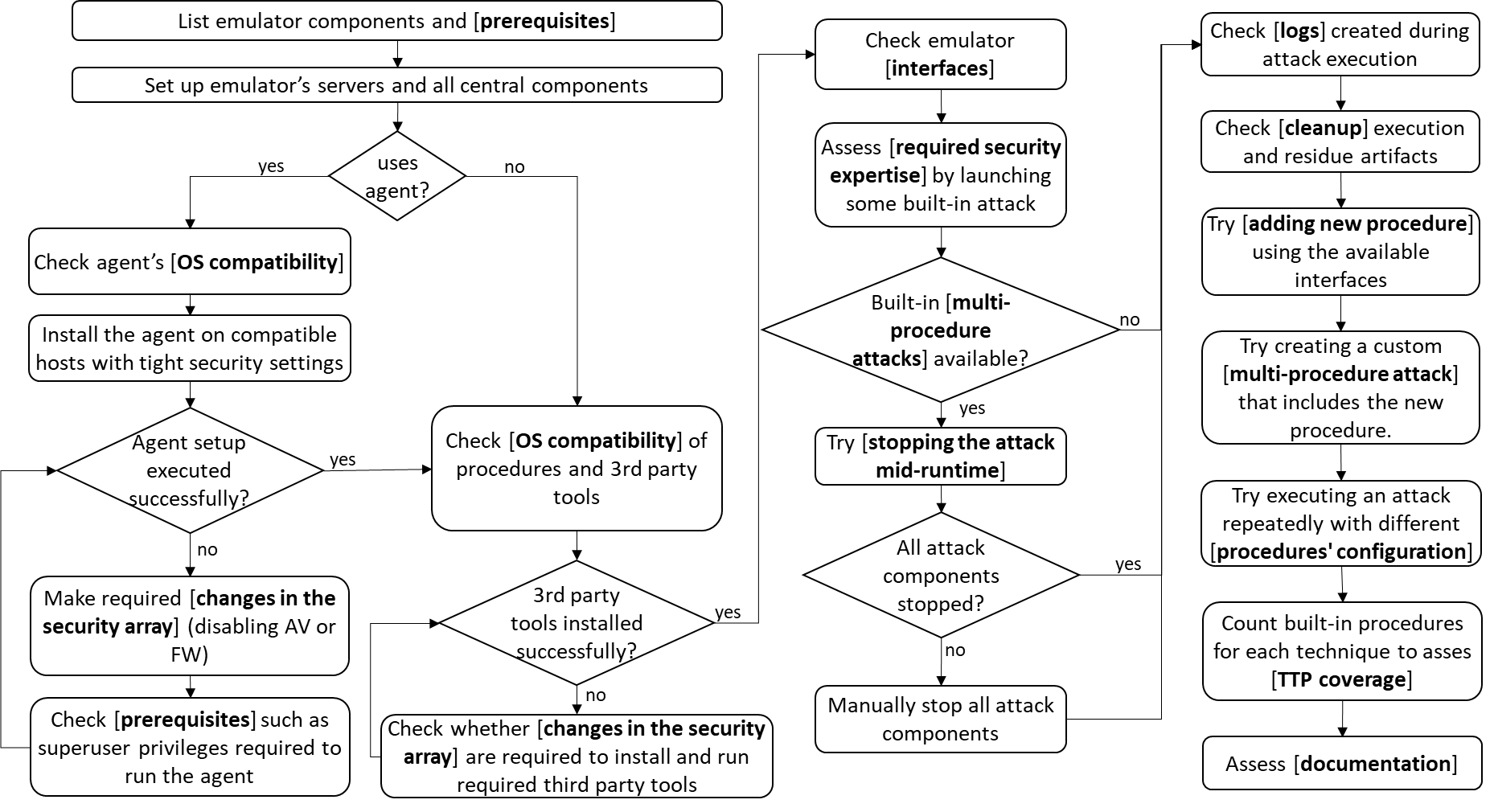}
    \caption{Threat emulator evaluation process}
    \label{fig:methodology}
\end{figure*}

\end{document}